\documentclass{article}
\usepackage{spconf,amsmath,graphicx}
\usepackage{threeparttable}
\usepackage{amssymb}
\usepackage{algorithm}
\usepackage{algorithmic}

\title{A Robust and Efficient Multi-Scale Seasonal-Trend Decomposition}
\name{Linxiao Yang\textsuperscript{\rm 1}, Qingsong Wen\textsuperscript{\rm 2}, Bo Yang\textsuperscript{\rm 1}, Liang Sun\textsuperscript{\rm 2}}
\address{
\textsuperscript{\rm 1}Machine Intelligence Technology, Alibaba Group, Hangzhou, China\\
\textsuperscript{\rm 2}Machine Intelligence Technology, Alibaba Group, Bellevue, USA
}

\begin{document}
%
\maketitle

\begin{abstract}
Many real-world time series exhibit multiple  seasonality with different lengths. The removal of seasonal components is crucial in numerous applications of time series, including forecasting and anomaly detection. However, many seasonal-trend decomposition algorithms suffer from high computational cost and require a large amount of data when multiple seasonal components exist, especially when the periodic length is long. In this paper, we propose a general and efficient multi-scale seasonal-trend decomposition algorithm for time series with multiple seasonality. We first down-sample the original time series onto a lower resolution, and then convert it to a time series with single seasonality. Thus, existing seasonal-trend decomposition algorithms can be applied directly to obtain the rough estimates of trend and the seasonal component corresponding to the longer periodic length. By considering the relationship between different resolutions, we formulate the recovery of different components on the high resolution as an optimization problem, which is solved efficiently by our alternative
direction multiplier method (ADMM) based algorithm. 
Our experimental results demonstrate the accurate decomposition results with significantly improved efficiency. 
\end{abstract}

\vspace{-3mm}
\section{Introduction}
\vspace{-3mm}

Recently, the explosive growth of the Internet of Things (IoT), Artificial Intelligence for IT Operations (AIOps) and many other applications leads to huge amounts of time series signals. Therefore, signal processing and mining for time series have received lots of research interests~\cite{isufi2019forecasting,wen2020robust,esling2012time}. Seasonality including multiple seasonality is commonly observed in time series data. For example, the traffic data typically exhibit the daily and weekly periodicities~\cite{gould2008forecasting}. Compared with single seasonality, multiple seasonality makes the seasonal-trend decomposition more challenging.

The seasonal-trend decomposition is an important procedure in the analysis of periodic time series as it is the basis for seasonal adjustment in many applications including forecasting and anomaly detection~\cite{Aminikhanghahi2017,hochenbaum2017automatic,laptev2015generic,theodosiou2011forecasting}. Recently, some seasonal-trend decomposition algorithms have been proposed, including MSTL~\cite{MSTL_hyndman2018package}, STR~\cite{STR_dokumentov2015str}, 
TBATS~\cite{TBATS_de2011forecasting} 
, and RobustSTL~\cite{RobustSTL_wen2018robuststl,wen2020fast}. These algorithms usually estimate different components iteratively using the original data. Thus, their computational cost is high, especially when the seasonality lengths become longer. Meanwhile, some multi-scale methods are proposed recently for time series storage, query processing~\cite{reeves2009managing}, and local pattern discovery~\cite{papadimitriou2006optimal} for improved efficiency.


In this paper, we propose a multi-scale seasonal-trend decomposition algorithm for time series with multiple seasonality. To the best of our knowledge, this work is the first one which applies the multi-scale approach in seasonal-trend decomposition. Let us consider the seasonal-trend decomposition of the traffic data again. Suppose we observe the traffic data every one minute. Note that we need sufficient data to estimate the seasonal components. For a single traffic time series, in order to accurately estimate both the daily and weekly seasonal components, the input time series should cover three weeks, i.e., $60\times 24\times 7 \times 3$ points. The resulting storage and computation cost is a burden in many applications as lots of time series are required to be processed in parallel. In our proposed framework, we model different components in different scales. Intuitively, the daily seasonal component is represented using the data recorded every minute, but the weekly seasonal component can be approximated using data on a lower resolution, e.g., hourly data. By considering different components at different resolutions, our seasonal-trend decomposition requires significantly less data and produce accurate components much more efficiently. For example, typically we can only use hourly data in the last three weeks and per-minute data in the last three days to accurately perform decomposition, which only requires $24\times 18 + 1440\times 3$ points (about $15.7\%$ of original data). 

Specifically, in our framework we first down-sample the time series onto a lower resolution. And then we compute the seasonal difference on the lower resolution using the shorter periodic length to reduce the shorter periodicity, based on which the rough estimates of trend and the other seasonal component can be obtained by applying existing seasonal-trend decomposition algorithms directly. Next we build the relationship between time series on different resolutions, and then decompose the time series on the high resolution by solving an optimization problem efficiently using our proposed ADMM \cite{boyd2011distributed} based algorithm. 






\section{Problem Statement}
Without loss of generality, in this paper we assume the input time series are with two different seasonal components. The method proposed in this paper can be extended to time series with more seasonal components by applying our algorithm repeatedly. Formally, we consider the time series $\{y_t\}$ can be decomposed as the sum of trend, seasonal and remainder components:
\begin{align}
    y_t = \tau_t + s_{d,t} + s_{w,t} + r_t, \quad t=-T_0+1,\cdots,0,
\end{align}
where $\tau_t$ and $r_t$ denote the trend and 
remainder components, respectively, $s_{d,t}$
and $s_{w,t}$ denote two seasonal components with period length $T_d$
and $T_w$ ($T_w>T_d$), respectively.
Here we denote current as $t=0$ for convenience. In the following, we call $s_{w_t}$, i.e., the seasonal component with longer periodic length $T_w$, ``the long seasonal component", and $s_{d,t}$ ``the short seasonal component". 
We assume that the trend is smooth locally and the remainder is composed of 
Gaussian noise $n_t$ and possible outliers $o_t$, i.e. $r_t=n_t+o_t$. 
$T_d$ and $T_w$ are assumed to be known, as these parameters can be generally
estimated accurately
using well-developed multiple periodicity detection and estimation methods, e.g. RobustPeriod~\cite{wen2020robustperiod}. 

To reduce the storage cost, in this paper we propose to store the time series data in a multi-scale manner. 
Specifically, for recent periods we store $y_t$
with high resolution, as we may need detailed information
to identify the anomalies or the change of trend; for the relatively older data, the low resolution is used for storage
to reduce the memory cost.
Mathematically,
with $\{\hat{y}_{t}^h\}$ and $\{\hat{y}_{t}^l\}$ denotes the stored 
high-resolution and low-resolution time series, respectively,
we have
\begin{align}
\hat{y}_{t}^h &= y_t \quad\quad\qquad\qquad\qquad -T_r<t\le 0\label{high-resolution}\\
\hat{y}_{t}^l &= \frac{1}{N}\sum_{i=(t-1)N+1}^{tN}y_i \qquad -T_0<tN\le-T_r\label{low-resolution}
\end{align}
where $T_r$ denotes the number of high-resolution data points
we stored, and $N$ denotes the down-sampling factor when store
the low-resolution data. Here we also assume
$T_r^l=T_r/N$, $T_d^l=T_d/N$ and $T_w^l=T_w/N$ are integer.


\vspace{-.3cm}
\section{Proposed Multi-Scale Decomposition}\label{section:decomposition}
In our multi-scale decomposition framework, we first down-sample the high-resolution data and obtain a full low-resolution time series. Then by exploiting the periodic nature of the short seasonality, we reduce the double periodic into a single periodic time series which is decomposed by applying existing decomposition algorithm. In this paper we select RobustSTL~\cite{RobustSTL_wen2018robuststl} due to its ability to handle abrupt trend change and seasonality shift. Finally, we build a relationship between the differentiation of the low-resolution time series and high-resolution time series,
and decompose the high-resolution time-series by
solving an optimization problem which utilizes the 
local smoothness of the trend. 

\vspace{-0.2cm}
\subsection{Decomposition of Low-Resolution Time Series}
\vspace{-.2cm}
In this subsection, we discuss how to decompose
the low-resolution time series using RobustSTL~\cite{RobustSTL_wen2018robuststl}. 
We augment $\hat{y}_{t}^l$ for $-T_r<t\le0$. According to 
(\ref{high-resolution}) and (\ref{low-resolution}),
we define
\vspace{-0.2cm}
\begin{align}
    \hat{y}_{t}^l
    &=\frac{1}{N}\sum_{i=(t-1)N+1}^{tN}\hat{y}_{i}^h, \quad -T_r<tN\le0.\label{low-resolution-aug} 
\end{align}
Let $\tau^l$, $s_{d,t}^l$, $s_{w,t}^l$ and $r_{t}^l$ denote the corresponding low-resolution of trend, two seasonal
and remainder components of the time series, respectively. Then
we have $\hat{y}_t^l = \tau^l_t + s_{d,t}^l + s_{w,t}^l + r_{t}^l$.



Note the down-sampled time series still contains two seasonal components. To further simplify the problem, we compute the seasonal difference of $\{\hat{y}_t^l\}$ using the short periodic length. 
Specifically, the difference of the time-series is 
\begin{align}
    g_t^l=\nabla_{T_d^l}\hat{y}_t^l=\hat{y}_{t+T_d^l}^l -\hat{y}_t^l
    = \nabla_{T_d^l}\tau^l_t + \nabla_{T_d^l}s_{w,t}^l + \tilde{r}_{t}^l\label{low-resolution-diff}
\end{align}
where
$\tilde{r}_{t}^l=\nabla_{T_d^l}s_{d,t}^l +\nabla_{T_d^l}r_{t}^l$. Here we use $\nabla_T$ to denote the difference of two points in a time series with a time difference of $T$ points. As $\tau_t$ is a local smooth signal with suddenly change,
it is easy to see that $\nabla_{T_d^l}\tau^l_t$ is also smooth locally.
Another observation is 
$\nabla_{T_d^l}s_{w,t}^l=s_{w,t+T_d^l}^l-s_{w,t}^l$ is periodic
and zero-mean time series
with period length equals to $T_w^l$ since $s_{w,t}^l$
is seasonal and period length equals to $T_w^l$.
Moreover, as the long seasonal component are assumed to change slowly across the nearby periods,
we conclude that $\tilde{r}_{t}^l$
can be modeled as a combination of Gaussian noise and outliers.
Based on the above discussion, we can assume that $g_t^l$ is a time series with single seasonality. Thus, $\nabla_{T_d^l}s_{w,t}^l$, $\nabla_{T_d^l}\tau^l_t$ and $\tilde{r}_{t}^l$ can be computed by using RobustSTL.


\vspace{-.25cm}
\subsection{Decomposition of High-Resolution 
Time Series}
\vspace{-.25cm}



In this subsection we discuss how to estimate different components on high resolution with the help of the decomposition results of low-resolution time series. To bridge the gap between the
high-resolution and low-resolution data, we compute the seasonal difference of the high-resolution of the data: 
\begin{align}
    g_t^h=\nabla_{T_d}\hat{y}_t^h=\hat{y}_{t+T_d}^h -\hat{y}_t^h= \nabla_{T_d}\tau^h_t + \nabla_{T_d}s_{w,t}^h + \tilde{r}_{t}^h\label{high-resolution-diff}
\end{align}
where
$
\tilde{r}_{t}^h=\nabla_{T_d}s_{d,t}^h +\nabla_{T_d}r_{t}^h.
$
Define $\boldsymbol{\tau}^l=[\tau_{0}^l,\dots,\tau_{-T_r^l+1}^l]^T$, $\boldsymbol{\tau}^h=[\tau_{0}^h,\dots,\tau_{-T_r+1}^h]^T$, $\boldsymbol{s}_{w}^l=[s_{w,0}^l,\dots,s_{w,-T_r^l+1}^l]^T$,
$\boldsymbol{s}_{w}^h=[s_{w,0}^h,\dots,s_{w,-T_r+1}^h]^T$,
$\boldsymbol{g}^h=[g_{0}^h,\dots,g_{-T_r+T_d+1}^h]^T$, and
$\boldsymbol{\tilde{r}}^h=[\tilde{r}_{0}^h,\dots,\tilde{r}_{-T_r+T_d+1}^h]^T$,
we have $\boldsymbol{\tau}^l=\boldsymbol{A}\boldsymbol{\tau}^h$,
$\boldsymbol{s}_{w}^l=\boldsymbol{A}\boldsymbol{s}_{w}^h$
and $\boldsymbol{g}^h=\boldsymbol{D}_{T_d}\boldsymbol{\tau}^h+\boldsymbol{D}_{T_d}\boldsymbol{s}_w^h+\boldsymbol{\tilde{r}}^h$
where $\boldsymbol{A}$ denotes the aggregation matrix, defined
as $\boldsymbol{A}=1/N(\boldsymbol{I}\otimes\boldsymbol{1}^T)$,
$\boldsymbol{I}\in\mathbb{R}^{T_r^l\times T_r^l}$ is an
identity matrix,
$\boldsymbol{1}\in\mathbb{R}^{N\times 1}$ is a vector
with all entries equal to $1$, and $\otimes$ denotes the Kronecker product,
$\boldsymbol{D}_{T_d}\in\mathbb{R}^{(T_r-T_d^l)\times T_r}$ 
denotes the difference matrix. The $i$th row of $\boldsymbol{D}_{T_d}$ is defined as a full zero vector with its $i$th and $(i+T_d^h)$th entries
equal to 1 and -1, respectively.
Let define $\nabla_{T_d^l}\boldsymbol{\tau}^l=[\nabla_{T_d^l}\tau_{0}^l, \nabla_{T_d^l}\tau_{-1}^l, \dots,\nabla_{T_d^l}\tau_{-T_r^l+1}^l]$ and define 
$\nabla_{T_d^l}\boldsymbol{s}_{w}^l=[\nabla_{T_d^l}s_{w,0}^l,\nabla_{T_d^l}s_{w,-1}^l,\dots,\nabla_{T_d^l}s_{w,-T_r^l+1}^l]$,
we have $\nabla_{T_d^l}\boldsymbol{\tau}^l=\boldsymbol{D}_{T_d^l}\boldsymbol{\tau}^l$ and $\nabla_{T_d^l}\boldsymbol{s}_{w}^l=\boldsymbol{D}_{T_d^l}\boldsymbol{s}_{w}^l$,
where $\boldsymbol{D}_{T_d^l}\in\mathbb{R}^{(T_r^l-T_d^l)\times T_r^l}$ 
denotes the difference matrix. The $i$th row of $\boldsymbol{D}_{T_d^l}$
defined as a full zero vector with its $i$th and $(i+T_d^l)$th entries 
equal to 1 and -1, respectively.
Then we propose to minimize following objective function to 
extract trend and the long period seasonality,
\begin{align}
    &\|\boldsymbol{g}^h-\boldsymbol{\widehat{D}}_{T_d}\boldsymbol{x}\|_1+\lambda_1\|\boldsymbol{z}-\boldsymbol{B}\boldsymbol{x}\|_2^2+\lambda_2\|\boldsymbol{\widehat{D}}\boldsymbol{x}\|_1
    +\lambda_3\|\boldsymbol{\widehat{D}}^2\boldsymbol{x}\|_1\label{loss-func}
\end{align}
where $\boldsymbol{\widehat{D}}_{T_d}=[\boldsymbol{D}_{T_d},\boldsymbol{D}_{T_d}]$, $\boldsymbol{x}=[(\boldsymbol{\tau}^h)^T,(\boldsymbol{s}_{w}^h)^T]^T$, $\boldsymbol{z}=[\overline{\nabla_{T_d^l}\boldsymbol{\tau}^l}^T,
\overline{\nabla_{T_d^l}\boldsymbol{s}_{w}^l}^T]^T$, $\boldsymbol{B}=\text{bdiag}(\boldsymbol{D}_{T_d^l}\boldsymbol{A},\boldsymbol{D}_{T_d^l}\boldsymbol{A})$, $\boldsymbol{\widehat{D}}=\text{bdiag}(\boldsymbol{D},\boldsymbol{D})$
and $\boldsymbol{\widehat{D}}^2=\text{bdiag}(\boldsymbol{D}^2,\boldsymbol{D}^2)$. Here $\text{bdiag}(\boldsymbol{A},\boldsymbol{B})$ is 
a block diagonal matrix with the two block are matrix $\boldsymbol{A}$ and $\boldsymbol{B}$,
respectively,
$\overline{\nabla_{T_d^l}\boldsymbol{\tau}^l}$
and $\overline{\nabla_{T_d^l}\boldsymbol{s}_{w}^l}$
are the values of $\nabla_{T_d^l}\boldsymbol{\tau}^l$
and $\nabla_{T_d^l}\boldsymbol{s}_{w}^l$ that obtained from
the decomposition of the low-resolution time series, respectively,
$\boldsymbol{D}\in\mathbb{R}^{(T_r-1)\times T_r}$ and $\boldsymbol{D}^2\in\mathbb{R}^{(T_r-2)\times (T_r-1)}$ denote the 
first and second order difference matrix, respectively. 
The $i$th row of $\boldsymbol{D}$ and $\boldsymbol{D}^2$
are defined as $[\boldsymbol{0}_i^T,1,-1,\boldsymbol{0}_{T_r-i-2}^T]$
and $[\boldsymbol{0}_i^T,-1,2,-1,\boldsymbol{0}_{T_r-i-3}^T]$,
where $\boldsymbol{0}_i$ denotes a zero vector of length $i$.
We note that the first term of $(\ref{loss-func})$ is the empirical loss, the second and third terms push the decomposition of the high-resolution
time series consistent with that of the low-resolution time series.
The last term of (\ref{loss-func}) forces the trend
and the long seasonal component to be smooth locally. We note that
in many practical applications the long seasonal component changes relatively slowly on the high resolution, thus it is reasonable to
assume its local smoothness.

\vspace{-.4cm}
\subsection{Efficient Implementation}

\begin{small}
\begin{algorithm}
	\renewcommand{\algorithmicrequire}{\textbf{Input:}}
	\renewcommand\algorithmicensure {\textbf{Output:}}
	\caption{Robust high-res time series Decomposition}
	\begin{algorithmic}[1]
		\REQUIRE High-resolution time series $\{\hat{y}_t^h\}$,
		$\overline{\nabla_{T_d^l}\boldsymbol{\tau}^l}$
and $\overline{\nabla_{T_d^l}\boldsymbol{s}_{w}^l}$ obtained from low-resolution time series decomposition.
		\ENSURE High-resolution $\tau_t$, $s_{d,t}$ and $s_{w,t}$.
		\STATE Denosing $\{\hat{y}_t^h\}$ using bilateral filter;
		\STATE Compute $g_t^h$ according to (\ref{high-resolution-diff});
		\STATE Initialize $\rho$, $\boldsymbol{x}$, $\boldsymbol{u}_1$, $\boldsymbol{u}_2$ and $\boldsymbol{u}_3$;
		\WHILE {not converge}
		\STATE Update $\boldsymbol{\bar{p}}$, $\boldsymbol{p}'$ and $\boldsymbol{p}''$ according to (\ref{aux-update});
		\STATE Update $\boldsymbol{x}$ using (\ref{x-update});
		\STATE Update $\boldsymbol{u}_1$,
		$\boldsymbol{u}_2$ and $\boldsymbol{u}_3$ according to (\ref{lagrange-update});
		\ENDWHILE
		\STATE Subtract $s_{d,t}$ using non-local seasonal filtering;
	\end{algorithmic}
	\label{algorithm-high-resolution}
\end{algorithm}
\end{small}
\vspace{-.2cm}


Here we apply the widely used ADMM to solve the resulting optimization problem as summarized in Algorithm 1. We first apply the variable splitting trick to split the smooth and non-smooth terms by introducing some auxiliary variables, and formulate the problem as
\begin{align}
    &\min\ \|\boldsymbol{g}^h-\boldsymbol{\bar{p}}\|_1+\lambda_1\|\boldsymbol{z}-\boldsymbol{{B}}\boldsymbol{x}\|_2^2
    +\lambda_2\|\boldsymbol{p}'\|_1
    +\lambda_3\|\boldsymbol{p}''\|_1\nonumber\\
    &\text{s.t.}\ \boldsymbol{\bar{p}} = \boldsymbol{\widehat{D}}_{T_d}\boldsymbol{x}\quad
    \boldsymbol{p}'=\boldsymbol{\widehat{D}}\boldsymbol{x}\quad
    \boldsymbol{p}''=\boldsymbol{\widehat{D}}^2\boldsymbol{x}
\end{align}
Consequently, the augmented Lagrange is
\begin{align*}\label{loss-func-lagrange}
    &\|\boldsymbol{g}^h-\boldsymbol{\bar{p}}\|_1
    +\boldsymbol{u}_1^T(\boldsymbol{\bar{p}}- \boldsymbol{\widehat{D}}_{T_d}\boldsymbol{x})+\rho/2\|\boldsymbol{\bar{p}}- \boldsymbol{\widehat{D}}_{T_d}\boldsymbol{x}\|_2^2\nonumber\\
    +&\lambda_1\|\boldsymbol{z}-\boldsymbol{{B}}\boldsymbol{x}\|_2^2
    +\lambda_2\|\boldsymbol{p}'\|_1
    +\boldsymbol{u}_2^T(\boldsymbol{p}'-\boldsymbol{\widehat{D}}\boldsymbol{x})
    +\rho/2\|\boldsymbol{p}'-\boldsymbol{\widehat{D}}\boldsymbol{x}\|_2^2\nonumber\\
    +&\lambda_3\|\boldsymbol{p}''\|_1+\boldsymbol{u}_3^T(\boldsymbol{p}''-\boldsymbol{\widehat{D}}^2\boldsymbol{x})+\rho/2\|\boldsymbol{p}''-\boldsymbol{\widehat{D}}^2\boldsymbol{x}\|_2^2,
\end{align*}
where $\rho$ is a pre-defined parameter.
Ignoring the terms independent with $\boldsymbol{x}$,
we can find the optimal $\boldsymbol{x}$ is given as
\begin{equation}\label{x-update}
\boldsymbol{x}=\boldsymbol{Q}^{-1}\boldsymbol{h}
\end{equation}
where $\boldsymbol{Q}=\rho\boldsymbol{\widehat{D}}_{T_d}^T\boldsymbol{\widehat{D}}_{T_d}+
2\lambda_1\boldsymbol{B}^T\boldsymbol{B}+
\rho\boldsymbol{\widehat{D}}^T\boldsymbol{\widehat{D}}+
\rho(\boldsymbol{\widehat{D}}^2)^T\boldsymbol{\widehat{D}}^2$
and $\boldsymbol{h}=\boldsymbol{\widehat{D}}_{T_d}^T(\boldsymbol{u}_1+\rho\boldsymbol{\bar{p}})+2\lambda_1\boldsymbol{B}^T\boldsymbol{z}+\boldsymbol{\widehat{D}}^T(\boldsymbol{u}_2+\rho\boldsymbol{p}')+(\boldsymbol{\widehat{D}}^2)^T(\boldsymbol{u}_3+\rho\boldsymbol{p}'')$. 
We note that the function with respect to the
auxiliary variables $\boldsymbol{\bar{p}}$,
$\boldsymbol{p}'$ and $\boldsymbol{p}''$ can be generally formulated
as the problem $\min_{\boldsymbol{x}} \lambda\|\boldsymbol{x}\|_1+\|\boldsymbol{y}-\boldsymbol{x}\|_2^2$,
and can be solved using proximal algorithms.
Then the updating rule of these auxiliary variables are given as
\begin{align}
\boldsymbol{\bar{p}}=&
\boldsymbol{g}^h-\mathcal{S}(\boldsymbol{g}^h-\boldsymbol{\widehat{D}}_{T_d}\boldsymbol{x}+1/\rho\boldsymbol{u}_1,1/\rho),
\nonumber\\
\boldsymbol{p}'=&
\mathcal{S}(\boldsymbol{\widehat{D}}\boldsymbol{x}-1/\rho\boldsymbol{u}_3,\lambda_2/\rho),\nonumber\\
\boldsymbol{p}''=&
\mathcal{S}(\boldsymbol{\widehat{D}}^2\boldsymbol{x}-1/\rho\boldsymbol{u}_3,\lambda_3/\rho)\label{aux-update}
\end{align}
where $\mathcal{S}(x,\rho)$ denotes the soft threshold operator
and is defined as $\mathcal{S}(x,\rho)=\text{sign}(x)\max(|x|-\rho,0)$. 
And the Lagrange multipliers can be updated using gradient ascend:
\begin{align}
\boldsymbol{u}_1^{(k+1)}&=\boldsymbol{u}_1^{(k)}+\rho(\boldsymbol{\bar{p}} - \boldsymbol{\widehat{D}}_{T_d}\boldsymbol{x}),\nonumber\\
\boldsymbol{u}_2^{(k+1)}&=\boldsymbol{u}_2^{(k)}+\rho(\boldsymbol{p}'-\boldsymbol{\widehat{D}}\boldsymbol{x}),\nonumber\\
\boldsymbol{u}_3^{(k+1)}&=\boldsymbol{u}_3^{(k)}+\rho(\boldsymbol{p}''-\boldsymbol{\widehat{D}}^2\boldsymbol{x}). \label{lagrange-update}
\end{align}

Note that ADMM is guaranteed to converge with a sufficient large $\rho$. To ensure fast converge, we initialize $\rho$ with a small value and increase it iteratively with a factor, say 1.15. 
Also the inverse of matrix $\boldsymbol{Q}$ can be computed offline, then the computational complexity of the proposed algorithm is dominated by the matrix-vector operation, e.g. $\boldsymbol{Q}^{-1}\boldsymbol{h}$, whose computational complexity is of order $\mathcal{O}(T_r^2)$.




\vspace{-.5cm}
\section{Experiments and Discussion}
\vspace{-.25cm}

In this section, we conduct experiments on both synthetic and public datasets to demonstrate the effectiveness of the proposed algorithm. 
Though out our experiments, we initialize $\rho$ with $10^{-5}$. We compare algorithm with STL~\cite{cleveland1990stl} and RobustSTL~\cite{RobustSTL_wen2018robuststl}. For those algorithms cannot handle multiple seasonality directly, we report the sum of all seasonal components as the final seasonal component.



\begin{figure}[t]
	\centering
    \includegraphics [width=3.8cm]{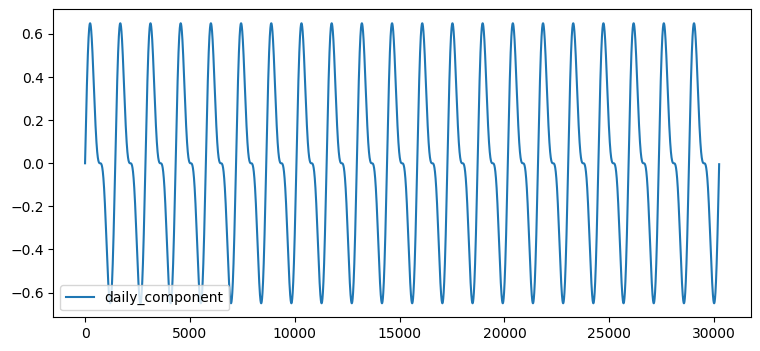}
    \includegraphics [width=3.8cm]{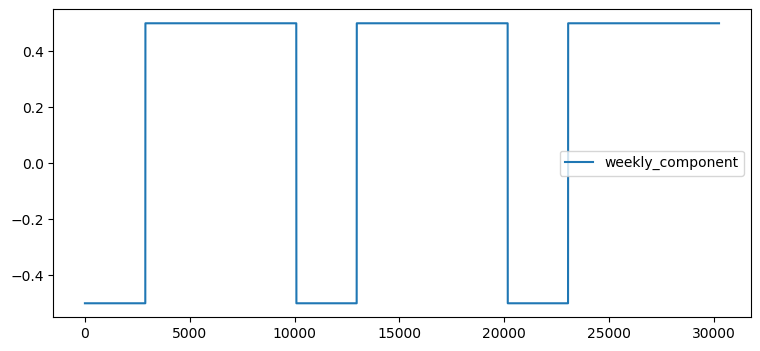}\\
    \includegraphics [width=3.8cm]{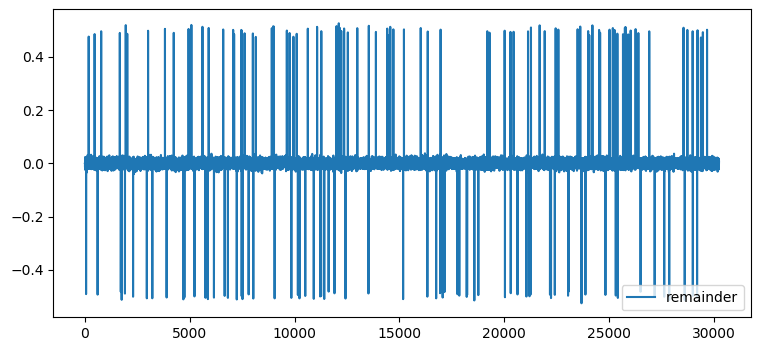}
    \includegraphics [width=3.8cm]{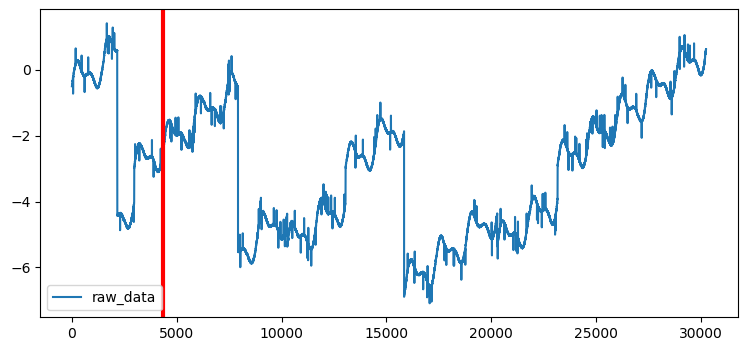}
	\caption{\footnotesize{Generated synthetic data. Top row from left 
	to right: generated seasonality $s_{d,t}$ and $s_{w,t}$,
	respectively. Bottom row from left to right:
	generated noise and raw time series. The red line in the bottom right splits the high-res and low-res data. }
	}
	\label{fig:syn-truth}
\end{figure}

\begin{figure}[t]
	\centering
    \includegraphics [width=2.7cm]{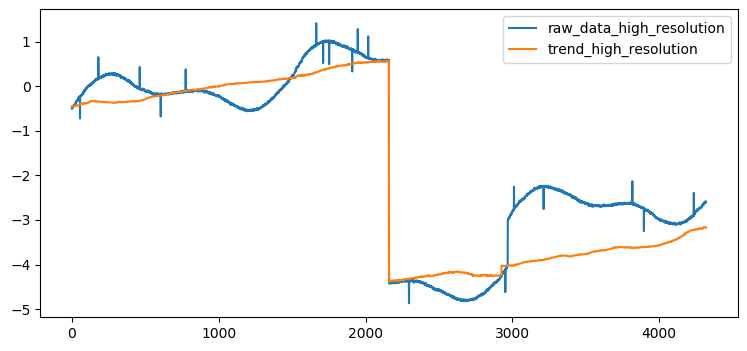}
    \includegraphics [width=2.7cm]{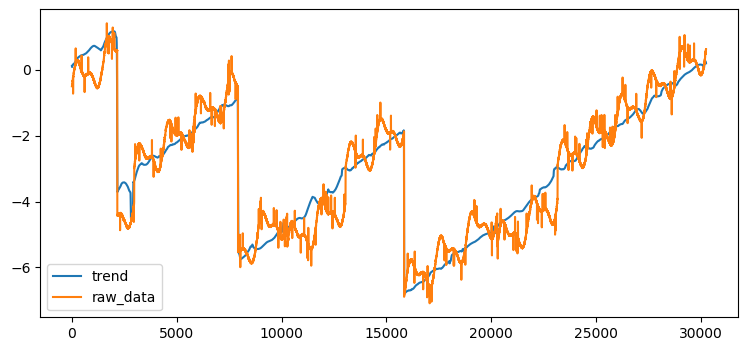}
    \includegraphics [width=2.7cm]{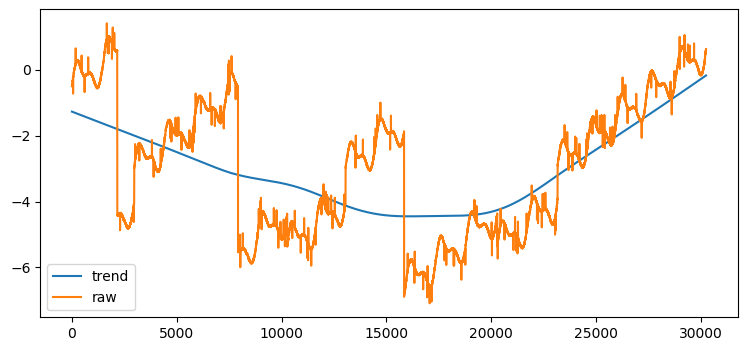}\\
    \includegraphics [width=2.7cm]{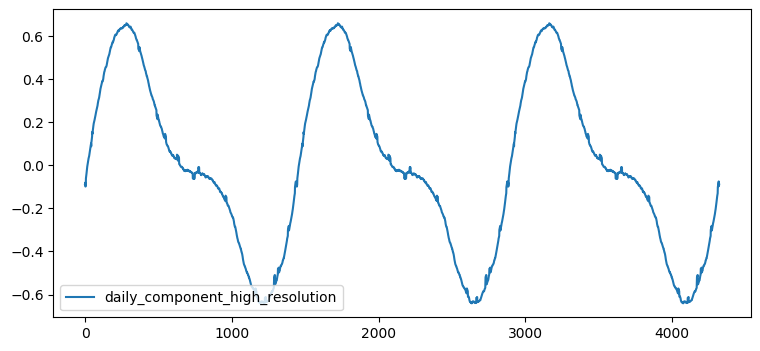}
    \includegraphics [width=2.7cm]{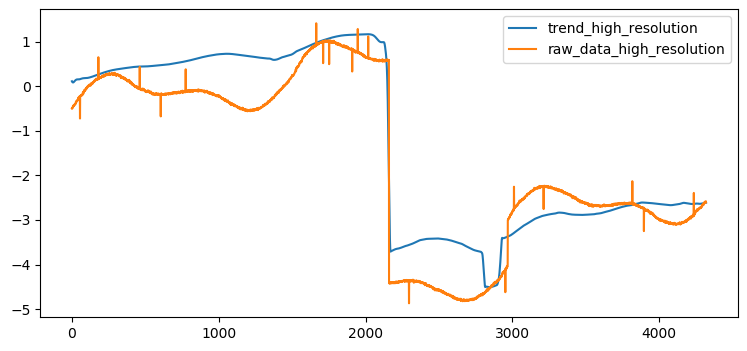}
    \includegraphics [width=2.7cm]{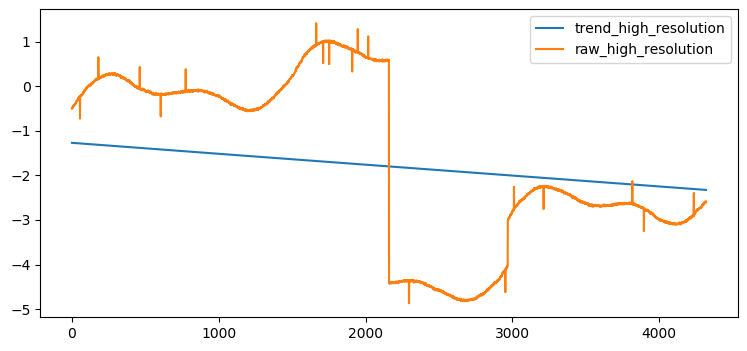}\\
    \includegraphics [width=2.7cm]{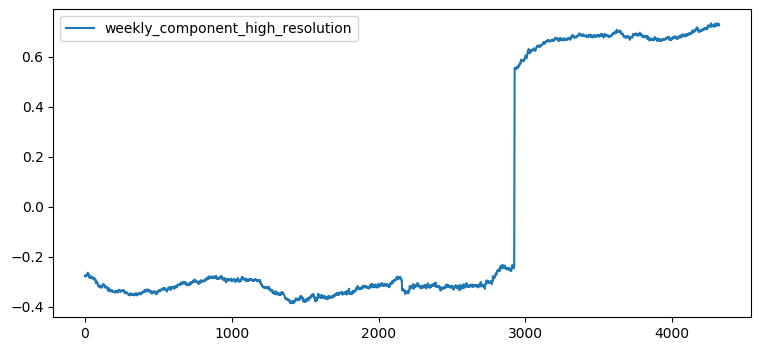}
    \includegraphics [width=2.7cm]{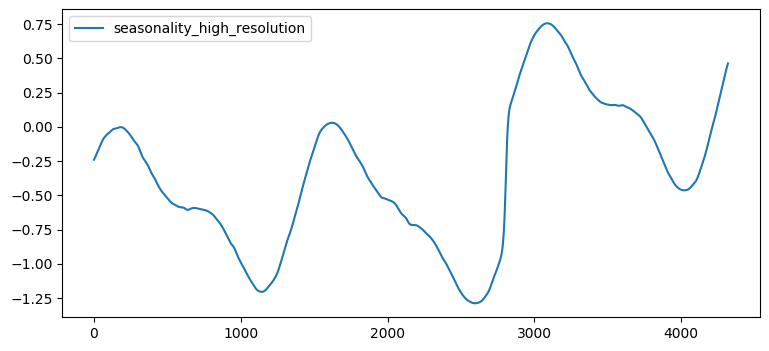}
    \includegraphics [width=2.7cm]{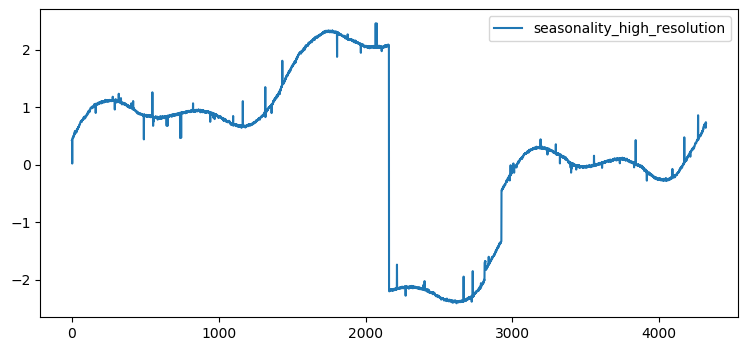}\\
    \includegraphics [width=2.7cm]{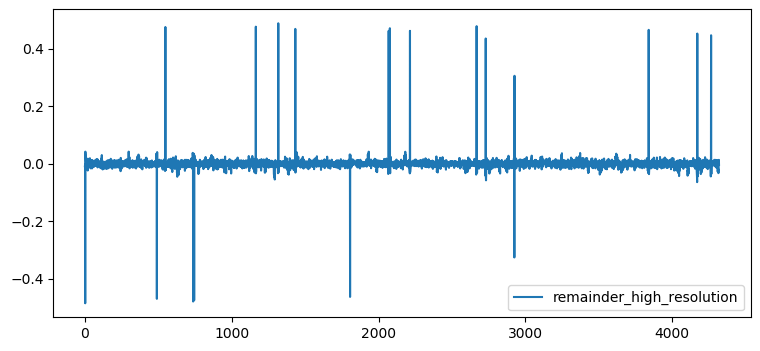}
    \includegraphics [width=2.7cm]{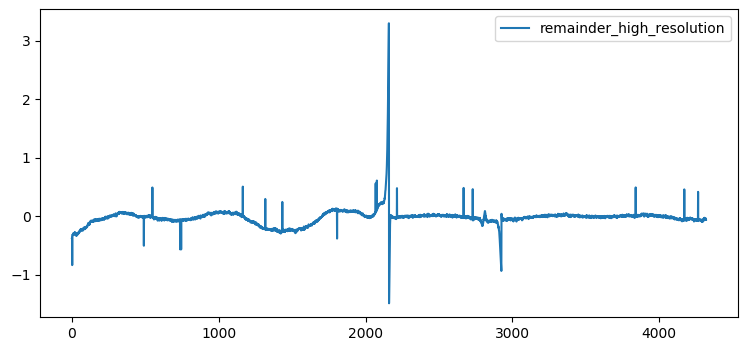}
    \includegraphics [width=2.7cm]{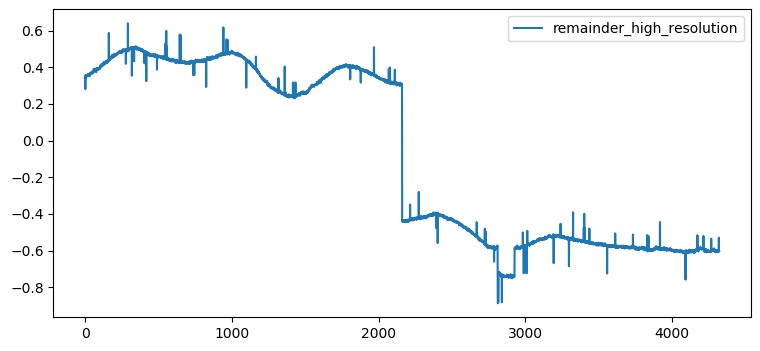}
	\caption{\footnotesize{Decomposition results on synthetic data. 
	From left to right are decomposition results of proposed method, RobustSTL and standard
	STL, respectively. First column (from top to bottom): decomposed trend, the short and long seasonal components, and remainder using proposed method, respectively.
	Second and third column (from top to bottom):  decomposed trends (low-res and high-res), seasonality and remainder, respectively. }
		}
	\label{fig:syn-decompose-result}
\end{figure}
\vspace{-0.4cm}

\vspace{-.45cm}
\subsection{Synthetic Data}
\vspace{-.15cm}

We generate a time series of length $30240$ containing two seasonality with periodic length of $1440$ and $10080$, respectively. We then add trend component with 3 abrupt changes,outliers as anomaly and white noise into the time series. Besides,
time warping is also performed to imitate the real-word data.
We keep the first $4320$ points as the high-resolution series and aggregate the rest as the low-resolution series. 
Here the aggregation factor $N$ is set to $60$.
Thus the high-resolution time series contains 
3 periods of the short seasonal component and 1 abrupt change of trend,
and the low-resolution time series is of length $432$, as illustrated in Fig.\ref{fig:syn-truth}.

Fig.\ref{fig:syn-decompose-result} summarizes the 
the decomposition results produced by three methods. Note that for STL and RobustSTL we use all the high-res data to achieve the decomposition. From Fig.\ref{fig:syn-decompose-result}, we observe that the proposed algorithm can decompose the short seasonal component accurately. It can also capture the abrupt trend change, outliers in remainder component. We also report the mean square error (MSE) of the  decomposed trend and seasonality, and the running time in seconds of all methods in Table \ref{table1}. It can be observed that proposed algorithm achieves the best trend and seasonal components with the least running time.

\vspace{-.2cm}
\begin{table}
    \small
	\centering
	\caption{MSE and running time on synthetic data}
	\begin{tabular}{c|c|c|c}
		\hline
		Method & Trend & Seasonality & Running Time (s) \\
		\hline
		\hline
		STL & 3.1012 & 1.7916 & 275.6 \\
		\hline
		RobustSTL  & 0.1090 & 0.0993 & 2138.23 \\
		\hline
		Proposed method & \textbf{0.0017}  & \textbf{0.0019}  & \textbf{17.4} \\
		\hline
	\end{tabular}
	\label{table1}
\end{table}

\vspace{-.25cm}
\subsection{Real Data}
\vspace{-.25cm}

\begin{figure}[t]
	\centering
    \includegraphics [width=8cm]{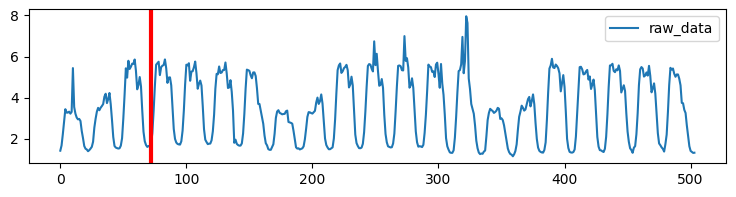}\\
\includegraphics [width=2.7cm]{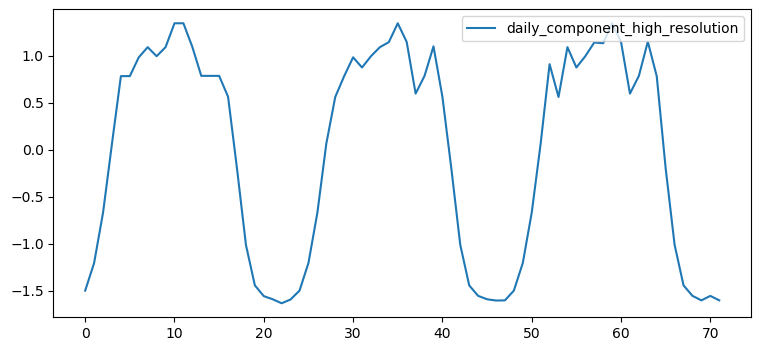}
\includegraphics [width=2.7cm]{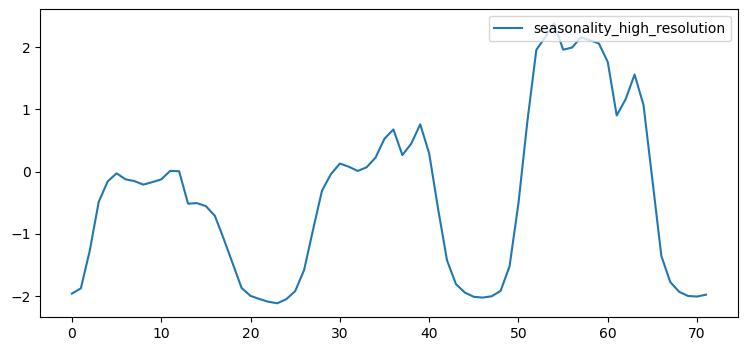}
\includegraphics [width=2.7cm]{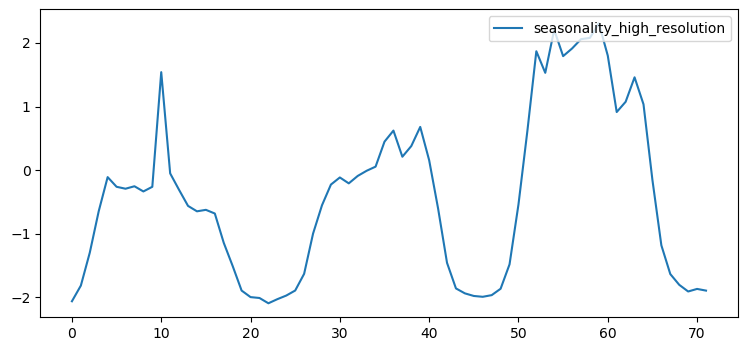}\\
\includegraphics [width=2.7cm]{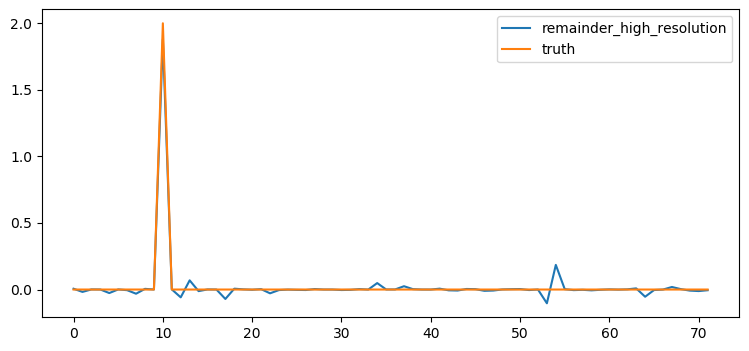}
\includegraphics [width=2.7cm]{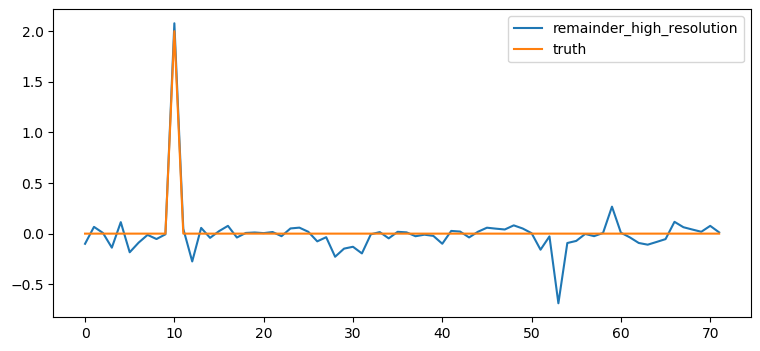}
\includegraphics [width=2.7cm]{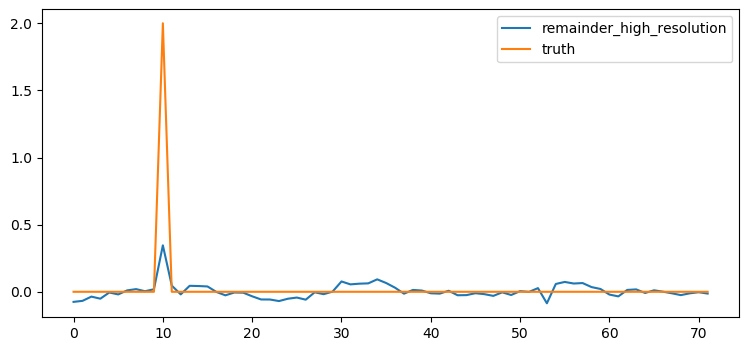}
	\caption{\footnotesize{Decomposition results on Yahoo! A1 dataset. From top to
	bottom are raw time series, decomposed seasonality and remainder by respective methods. The red line of the top sub-image splits the high-res and low-res data. From left to right of second and third rows are results obtained using proposed method and RobustSTL and STL, respectively.}
	}\vspace{-.25cm}
	\label{fig:real-data}
\end{figure}

We next compare different algorithms on the public Yahoo A1 dataset~\footnote{https://webscope.sandbox.yahoo.com/catalog.php?datatype=s\&did=70}. It is a collection of real production traffic to some of the Yahoo! properties with a total length of $504$ and two seasonal components (periodic lengths are 24 and 168). 
We set first $72$ data points as high-resolution time series and aggregate the rest with a factor 4 as low-resolution data. We inject one outlier as anomaly to show the robustness of the proposed
method. Three algorithms are compared in the same setting as on the synthetic data.


Fig.\ref{fig:real-data} summarizes the decomposition results. Here we only plot the short seasonal component learned by the proposed method. 
From Fig.\ref{fig:real-data} we can observe that both the proposed method and RobustSTL are able to learn a decent seasonality and capture the outlier in the remainder, while STL fails. We also observe that the remainder decomposed by the proposed method involves fewer noise than RobustSTL.

\vspace{-0.3cm}
\section{Conclusion}\label{sec:conc}
\vspace{-0.3cm}

In this paper we propose a robust and efficient seasonal-trend decomposition algorithm for time series with multiple seasonality using the multi-scale approach. It can achieve accurate decomposition with significant reduced storage and computation. In the future we plan to apply it in long-term and short-term forecasting by further utilizing the decomposed components at different resolutions.


\bibliographystyle{IEEEbib}
\bibliography{5_RVFbibfile}

\end{document}